\documentclass[%
 reprint,
 amsmath,amssymb,
 aps,
 prl,
]{revtex4-2}

\usepackage{graphicx}
\usepackage{dcolumn}
\usepackage{bm}

\usepackage{lipsum}
\usepackage{amsmath}
\usepackage{graphicx}
\usepackage{color}
\usepackage{tikz}
\usepackage{verbatim}
\usepackage{hyperref}
\usepackage{amssymb}
\usepackage{bm}
\usepackage[justification=raggedright]{caption}
\usepackage{subcaption}
\usepackage{graphicx}
\usepackage{subcaption}
\usepackage[export]{adjustbox}
\usepackage{wrapfig}
\usepackage{tabularx}
\usepackage[capitalise]{cleveref}
\usepackage{afterpage}  

\newcommand{\rom}[1]{\textup{\uppercase\expandafter{\romannumeral#1}}}

\allowdisplaybreaks

\begin{document}

\title{Eff-ACT-ive Starobinsky pre-inflation}

\author{Abhijith Ajith}
    \email{abhijith.ajith.421997@gmail.com}

\affiliation{Indian Institute of Science Education and Research Bhopal, Bhopal, Madhya Pradesh - 462066, India}

\author{Hardik Jitendra Kuralkar}
     \email{kuralkarhardik@gmail.com}

\affiliation{Indian Institute of Science Education and Research Bhopal, Bhopal, Madhya Pradesh - 462066, India}

\author{Sukanta Panda}
     \email{sukanta@iiserb.ac.in}

\affiliation{Indian Institute of Science Education and Research Bhopal, Bhopal, Madhya Pradesh - 462066, India}

\author{Archit Vidyarthi}
    \email{architmedes@gmail.com}
       
\affiliation{Indian Institute of Science Education and Research Bhopal, Bhopal, Madhya Pradesh - 462066, India}
\affiliation{Indian Institute of Technology Bombay, Powai, Mumbai, Maharashtra - 400076, India}
\date{\today}

\begin{abstract}
We consider quantum corrections to a recently obtained perturbative form of Starobinsky model to extract information about the initial conditions of the universe leading to cosmological inflation. Integrating out graviton modes, we find higher-derivative instabilities that are shown to decay into scalarons, causing an effective kinetic-domination stage which is shown to lead naturally to inflation without the need for fine-tuning of initial inflaton amplitude. We find a perturbative upper bound on scalaron magnitude that matches Planck constraints on inflaton energy density near the pivot scale. This modified history also affects observables and resolves other anomalies such as the low-$\ell$ power deficit in the TT spectrum as well as the disfavorment of Starobinsky model based on updated predictions of scalar spectral index accounting for data from Atacama Cosmology Telescope (ACT).

\end{abstract}

\maketitle
Inflationary models of the early universe \cite{Guth:1980zm,Starobinsky:1980te,Albrecht:1982wi,Linde:1983gd,Golovnev:2008cf,Aashish:2021gdf,Ajith:2022wia,Ajith:2025rty} provide a compelling framework for explaining the observed large-scale homogeneity, isotropy, and the generation of superhorizon primordial perturbations observed in the Cosmic Microwave Background (CMB) anisotropy \cite{Mather:1990tfx}. But the paradigm still faces several challenges. One critical issue is the fine-tuning problem: viable inflation consistent with observational constraints can be achieved for only a small portion of the parameter space for which there exist very few theoretical frameworks. Another problem is the power deficit observed in large angular temperature correlations compared to predictions for $\Lambda$CDM cosmology \cite{Mather:1990tfx,Planck:2018jri}. More recently, results from ACT showed an enhancement in the scalar spectral index to $0.9743\pm0.0034$ (for Planck+ACT+LB data) which has ruled out several well-motivated inflation models.

Despite the success of Starobinsky inflation model \cite{Starobinsky:1980te} in light of experimental constraints on temperature and polarization anisotropies \cite{Planck:2018jri}, it was largely disfavored by the refined predictions of scalar spectral index by ACT. Since Starobinsky model had proven to be a good fit with most available data, we expect that quantum effects near the inflationary scales might help restore its glory, similar to the case of Higgs-like inflation \cite{Gialamas:2025kef}.

We begin our analysis with a perturbative version of Starobinsky action obtained in our previous work \cite{Mohanty:2025jeu},
\begin{align}\label{eq:quadratic-action-zeta-quad}
    S&=\int d^4x \left[\left(1+\sqrt{\frac{2}{3}}\frac{\zeta}{M_P}\right)\left(-\partial_\mu h^{\mu\nu}\partial_\nu h+\frac{1}{2}\partial_\mu h\partial^\mu h\right.\right.\nonumber\\
    &\left.\left.-\frac{1}{2}\partial_\rho h^{\mu\nu}\partial^\rho h_{\mu\nu}+\partial_\mu h^{\mu\nu}\partial^\rho h_{\rho\nu}\right)-\frac{1}{2}\partial_\mu\zeta\partial^\mu\zeta-\frac{M_\zeta^2}{2}\zeta^2\right.\nonumber\\
    &\left.+\frac{2}{3}\frac{\zeta}{M_P}(\partial_\mu h^{\mu\nu}\partial_\nu\zeta-\partial_\mu h\partial^\mu\zeta)+\frac{1}{\sqrt{6}}\frac{\zeta}{M_P}\partial_\mu\zeta\partial^\mu\zeta\right],
\end{align}
where $\zeta$ is the scalaron with mass $M_\zeta\approx1.3\times10^{-5}M_P$ based on Planck 2018 results \cite{Planck:2018jri}, where $M_P$ is the reduced Planck mass. This form was obtained under the conditions of flat spacetime background and small scalaron magnitude. Note that additional interactions from higher order expansion of $R$ only modify the coupling coefficients which doesn't affect the overall dynamics drastically, so we can safely restrict ourselves to \eqref{eq:quadratic-action-zeta-quad}.

We also assume a flat FLRW background for our pre-inflationary analysis. On sub-horizon scales, the background can be considered Minkowskian and since both $h_{\mu\nu}$ and $\zeta$ are canonical, we can quantize the theory without any issues. The inflationary era in Starobinsky's model is characterized by $H\sim M_\zeta$. To ensure that leading order quantum corrections suffice to describe quantum deviations from classical behavior, we maintain $H\lesssim 10^{-2}M_P$ throughout this work. As we shall see, quantum corrections indeed play a big role in mitigating all three of the aforementioned issues plaguing inflation and the Starobinsky model.

We obtain leading order corrections to the $\zeta$ part of the action by integrating out graviton modes. From \eqref{eq:quadratic-action-zeta-quad}, only the terms that are quadratic in $h_{\mu\nu}$ contribute to the one-loop effective action. Ignoring background curvature, we find that the effective scalaron action becomes,
\begin{align}
    S_{\rm eff}[\zeta]\approx\int d^4x &\left[-\frac{1}{2}\partial_\mu\zeta\partial^\mu\zeta-\frac{1}{2}M_\zeta^2\zeta^2+\frac{1}{\sqrt{6}}\frac{\zeta}{M_P}\partial_\mu\zeta\partial^\mu\zeta\right.\nonumber\\
    &\left.+\frac{\beta}{M_P^2}\Box\zeta\Box\zeta\right],
\end{align}
up to vacuum corrections, where $\beta$ is a dimensionless parameter and the higher-derivative term arises from $\zeta h^2$-type nonlinear kinetic interactions in the background action \eqref{eq:quadratic-action-zeta-quad}, which are treated as effective mass terms for the gravitons. Higher-derivative terms in the effective action generally suggest missing interactions with external sources. Since we’ve integrated out quantized graviton perturbations here, these represent missing graviton exchange interactions. The higher-derivative term introduces an unstable mode, as per the Ostrogradsky theorem \cite{Ostrogradsky:1850fid,Joshi:2023otx}. This unstable mode can decay into stable modes through available couplings, similar to Lee-Wick fields \cite{Grinstein:2007mp}. Generally, decay of Lee-Wick fields into stable fields is associated with violation of micro-causality. However, since the higher-derivative term appears here due to us having integrated out graviton modes, the supposed violations are not actual violations. 

For the purpose of this analysis, we impose a constraint $\beta\ll M_P^2/M_\zeta^2$. Small $\beta$ (or equivalently large mass of unstable, decaying particle $M_\zeta\ll M_P/\sqrt{\beta}$) is consistent with quantum gravitational effects becoming relevant close to the Planck scale. Rewriting the action in terms of stable field $\hat{\zeta}$ and unstable field $\Tilde{\zeta}$ (see \cite{Grinstein:2007mp}), it can be shown that the kinetic-interaction term yields the following decay vertices,
\begin{equation}
    -\sqrt{\frac{2}{3}}\frac{\hat{\zeta}}{M_P}\partial_\mu\hat{\zeta}\partial^\mu\Tilde{\zeta}-\frac{1}{\sqrt{6}}\frac{\Tilde{\zeta}}{M_P}\partial_\mu\hat{\zeta}\partial^\mu\hat{\zeta},
\end{equation}
with $M_P/\sqrt{\beta}$ being the mass of decaying particle. In the present case, the decay width is $\Gamma\sim M_P\beta^{-3/2}$. Imposing instantaneous decay requires $\Gamma\gg H$ which is ensured by constraining $\beta\lesssim\mathcal{O}(10^1)$ for $H\sim 10^{-2}M_P$. This constraint on $\beta$ combined with the small scalaron mass implies that the decay products possess large kinetic energy, i.e. kinetic part of scalaron energy density dominates in this regime. Thus, there must be a kinetic-dominated regime before inflation. Physically, this energy transfer is sourced by quantized gravity fluctuations in the pre-inflationary universe.  


We assume the unstable field $\Tilde{\zeta}$ has decayed into the stable field $\hat{\zeta}$, and work with the background action. For typographical ease, we revert to the earlier notation with $\hat{\zeta}\to\zeta$. Generalizing \eqref{eq:quadratic-action-zeta-quad} to a curved background,
\begin{align}\label{eq:kination-action-nonminimal}
    S\approx\int d^4x \sqrt{-g}&\left[\frac{M_P^2}{2}R-\frac{1}{2}\left(1-\sqrt{\frac{2}{3}}\frac{\zeta}{M_P}\right)\partial_\mu\zeta\partial^\mu\zeta\right.\nonumber\\
    &\left.-\frac{1}{2}M_\zeta^2\zeta^2\right],
\end{align}
where the non-minimal coupling can be shown to be absent up to $\mathcal{O}(\zeta^2/M_P^2)$ due to corrections from the integration measure function $\sqrt{-g}$ \cite{Mohanty:2025jeu}. The modification to the $\zeta$ kinetic term can be neglected since we are working in the $\zeta\ll M_P$ limit consistent with the kination era. In an FLRW spacetime background, the momentum space counterpart $\zeta(k)$ follows the equation of motion,
\begin{equation}\label{eq:zeta-mode-eom-kination}
    \Ddot{\Bar{\zeta}}+3H\Dot{\Bar{\zeta}}+\frac{k^2}{a^2}\Bar{\zeta}\approx0,
\end{equation}
where we have ignored the mass term considering that we're working in the limit $M_P\gg H\gg M_\zeta$. During kination, total energy density $\approx\Dot{\zeta}^2/2$ and equation of state parameter $w\approx1$ such that $H\sim1/(3t)$ and rate of expansion $a\propto t^{1/3}$, where $t$ represents the cosmic time. The general solution to \eqref{eq:zeta-mode-eom-kination} is given as,
\begin{equation}\label{eq::zeta-modes-kination}
    \Bar{\zeta}(\Vec{k},t)=c_1J_0\left(\frac{k}{a}t\right)+c_2Y_0\left(\frac{k}{a}t\right),
\end{equation}
where $J_n(x)$ and $Y_n(x)$ are Bessel's functions of first and second kind, respectively. 
Even if $\zeta$ is assumed to be non-homogeneous at the beginning of kination, since $a^{-2}\Vec{\nabla}^2\propto t^{-2/3}$, the spatial derivative term decays with time and $\zeta$ is eventually homogenized. This implies that the background value of $\zeta$ increases with time,
\begin{equation}\label{eq:kination-solution-zeta}
    \zeta_{f}=\zeta_{i}+A\ln{\left(\frac{H_{i}}{H_{f}}\right)}=\zeta_{i}-A\ln{\left(\frac{H_{f}}{H_{i}}\right)},
\end{equation}
for some integration constant $A$ which quantifies the increase in scalaron magnitude with decreasing $H$. Assuming a natural initial condition $\zeta_i\ll M_P$, \eqref{eq:kination-solution-zeta} provides a logarithmic relation between the integration constant $A$ and $H_i$ at the beginning of kination. Remember that $H_i$ is fixed by the assumption that the effective theory is valid and that the leading order quantum corrections suffice to describe the modified dynamics. As $\zeta$ increases, the effective kinetic term rapidly diminishes in magnitude. It can be shown through stability analysis that inflation follows directly from such a kination era. As $\zeta\to\sqrt{3/2}M_P$, the potential energy density attains its highest value $0.75M_\zeta^2M_P^2$. The condition $\zeta\lesssim\sqrt{3/2}M_P$ poses a natural positivity bound on $\zeta$ during kination. Note that even though it appears that we are at risk of encountering ghosts beyond this cut-off, the upper limit simply marks the failure of the perturbative regime. Given the fact that the estimated total energy density around pivot exit ($k_*=0.05 \text{Mpc}^{-1}$) as per Planck data $\approx0.73M_\zeta^2M_P^2$ is quite close to this bound, we can effectively proceed knowing that the analysis presented here is valid up to the start of inflation. Approximating $H_i\approx10^{-2}M_P$ and $H_f\approx6.5\times10^{-6}M_P$ (using $0.75M_\zeta^2M_P^2\approx3M_P^2H_f^2$), we can fix $A\approx0.17M_P$. It can be shown that for $H_f<H_i\lesssim10^{-2}M_P$, $0<N_{\rm kin}\lesssim2.4$. 

\begin{figure}
    \centering
    \includegraphics[width=\linewidth]{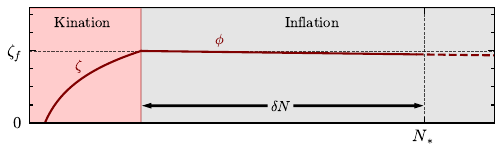}
    \caption{Evolution of field magnitude across the kination and inflation eras. Note that the 0 reference point is only for field magnitude and not $N$.}
    \label{fig:schematic}
\end{figure}


Given the failure of perturbation theory near $\zeta\to\sqrt{3/2}M_P$, it is clear that as we enter the inflationary era, we need to embrace the Einstein frame Starobinsky model with potential,
\begin{equation}\label{eq:starobinsky-einstein-action}
    \frac{3}{4}M_\zeta^2M_P^2\left(1-e^{-\sqrt{\frac{2}{3}}\frac{\phi}{M_P}}\right)^2,
\end{equation}
which cannot be obtained by a field-transformation between $\zeta$ and $\phi$ for two reasons: the perturbative expansion used to obtain \eqref{eq:kination-action-nonminimal}, as well as graviton modes having been integrated out of the system. While one may obtain a smoother transition across the kination-inflation boundary beyond perturbation theory, we use the assumption that the evolution from kination to the start of inflation is governed by \eqref{eq:kination-action-nonminimal} while dynamics from the onset of inflation are governed by the potential \eqref{eq:starobinsky-einstein-action}. This \textit{stitching} is motivated by the fact that for $\phi\ll M_P$, \eqref{eq:starobinsky-einstein-action} reduces to $M_\zeta^2\phi^2/2$ and for $\zeta\ll M_P$, the kinetic coupling coupling in \eqref{eq:kination-action-nonminimal} vanishes and we recover the Einstein frame Starobinsky model for small field magnitude. Thus, $\phi$ can be identified with $\zeta$. For $\phi\gg M_P$ we again see the potentials from the two eras match: $3M_\zeta^2M_P^2/4$. The evolution of $\zeta$ and $\phi$ during kination and inflation eras is shown schematically in Fig. \ref{fig:schematic}.

We now quantify the effects of this pre-inflationary kination era on the scalar power spectrum. We assume that the modes that exit the horizon at the pivot scale originated deep inside the horizon during the kination era. The spatial gradients have been shown to have decayed sufficiently by the end of kination so the fluid approximation is considered valid and the system is approximately free from anisotropic stress. Perturbations in the two eras are then matched across the transition boundary.

The modified scalar power spectrum is found to be,
 \begin{equation}\label{eq:power-spectrum-modified}
     \mathcal{P}_s(k)=\mathcal{P}^{\rm BD}_s(k)T(k)=\mathcal{A}_s\left(\frac{k}{k_*}\right)^{n_s-1}T(k),
 \end{equation}
 where the superscript BD helps identify the Bunch-Davies power spectrum with spectral index $n_s$, and the transfer function,
\begin{align}\label{eq:transfer-function}
   &T(k)= \frac{\pi}{2 k|\tau|}\Bigg[(-k\tau)^2\left(J_{\chi}^2\left(k\tau\right)+J_{\chi+1}^2\left(k\tau\right)\right) \nonumber\\
   & +2 k\tau\left(\chi+\frac{1}{2}\right) J_{\chi}\left(k\tau\right) J_{\chi+1}\left(k\tau\right)+\left(\chi+\frac{1}{2}\right)^2 J_{\chi}^2\left(k\tau\right)\Bigg]_{\tau_f},
\end{align}
shows decaying-oscillatory behavior and quantifies the modification due to pre-inflationary kination. A similar expression was obtained in \cite{Das:2014ffa} for pre-inflationary radiation domination era. Assuming that it takes $\delta N$ e-folds for the deep horizon modes at the start of inflation to exit the horizon,
\begin{equation}
    \tau_f=\tau_*e^{\delta N}\frac{H_*}{H_f},
\end{equation}
where the subscript $*$ refers to values at pivot exit and $f$ refers to values at transition boundary. Therefore, $T(k)$ can be shown to depend on $\delta N$. The corresponding modification to the scalar spectral index can be calculated as,
\begin{equation}\label{eq:deltans}
    \Delta n_s\equiv n^{\rm eff}_s-n_s^{\rm BD}=\frac{d\ln{T(k)}}{d\ln k}.
\end{equation}

The primordial power spectrum, therefore, gets modified through the parameter $\delta N$. We try to constrain this parameter from CMB data. For this, we implement our inflationary power spectrum in a modified version of the Boltzmann solver code \texttt{CAMB} \cite{Lewis:1999bs,2012JCAP...04..027H}, and perform Markov-Chain-Monte Carlo (MCMC) simulations using the publicly available tool \texttt{COBAYA} \cite{Torrado:2020dgo,2019ascl.soft10019T}. Apart from the parameter $\delta N$ characterizing the modified power spectrum, we vary other standard parameters, namely, the physical densities of baryons ($\Omega_b  h^2$) and dark matter ($\Omega_{c}h^2$), angular size of the sound horizon at recombination ($\theta_s$), the amplitude ($\mathcal{A}_s$) of the primordial power spectrum at pivot scale ($k_*=0.05$Mpc$^{-1}$), and the optical depth of reionization ($\tau_{\rm reio}$) in our MCMC analysis. For Starobinsky inflation, the scalar spectral index,
\begin{equation}\label{eq:nsbd}
    n_s^{\rm BD}\approx 1- \frac{2}{N_*}=1-\frac{2}{(N_{\rm tot}-\delta N)},
\end{equation}
where $N_{\rm tot}$ is the total number of inflationary e-folds (constrained by horizon and flatness problems) and $N_*$ is the number of e-folds from the end of inflation to pivot exit (constrained via reheating). Hubble radius increases during a pre-inflationary kination era, bringing in modes from outside the horizon. Thus, modes in causal contact today may not have been so during the kination phase, reducing the number of inflationary e-folds necessary to solve the horizon problem \cite{riotto2002inflation}. We have chosen a conservative $N_{\rm tot}=60$ for our analysis. Standard Starobinsky reheating dynamics suggests $N_*=54$ \cite{Planck:2018jri}. However, our previous work studying gravitational wave production based on modified reheating dynamics \cite{Mohanty:2025jeu} suggests a slightly lower value $N_*\approx51$, while another recent analysis find that for agreement with ACT data $N_*\approx 67$ \cite{Zharov:2025zjg}. $\delta N$ can therefore assume a large range of values. Due to this uncertainty, we proceed without fixing $N_*$ and try to find the best-fit value of $\delta N$ from CMB which will indirectly give us an approximate value of $N_*$ as well. The corresponding prior intervals for each of the parameters used in our analysis are given in \cref{prior}.
\begin{table}[h!]\label{data}
    \centering
    \begin{tabular}{ccc}
        \hline
        \hline
    \hspace{1em}\textbf{Parameters}\hspace{1em} & \hspace{1em}\textbf{Prior Interval}\hspace{1em} &\\
        \hline 
        
        $\Omega_b  h^2$ & [0.017, 0.027] &\\
        $\Omega_{c}h^2$ & [0.09, 0.15] &\\
        $ {\rm{ln}}(10^{10} \mathcal{A}_s)$ & [2.6, 3.5] &\\
        $\tau_{\rm reio} $ & [0, 0.1] &\\        
        $100\theta_s$ & [1.03, 1.05]&\\
       $\delta N$ & [0, 20]&\\
        \hline
    \end{tabular}
    \caption{Prior intervals for cosmological parameters}
    \label{prior}
\end{table}

In order to constrain these parameters, we use the corresponding likelihoods employed in the ACT DR6 analysis \cite{AtacamaCosmologyTelescope:2025blo}. We consider combinations of the high-$\ell$ TTTEEE ACT DR6 \texttt{lite} likelihood \cite{AtacamaCosmologyTelescope:2025blo}, low-$\ell$ TT (\texttt{Commander}) and EE (\texttt{Sroll2}) likelihoods \cite{Planck:2019nip,Delouis:2019bub,Pagano:2019tci}, the ACT DR6 lensing likelihood \cite{ACT:2023kun} and BAO from DESI DR1 \cite{DESI:2024mwx}. We refer to this combination as P-ACT-LB. When combining Planck and ACT DR6 we follow the ACT collaboration’s recommendation and impose a cut in the multipoles. The convergence of chains is ensured by having the Gelman-Rubin criterion $|R-1| \leq 0.01$. We utilize  \texttt{GetDist} \cite{Lewis:2019xzd} and \texttt{BOBYQA} \cite{2018arXiv180400154C,2018arXiv181211343C} to analyze the chains and maximize the likelihood posterior, respectively. The $68\%$ constraints on cosmological parameters along with their best-fit values are reported in \cref{tab:cons_table}.
\begin{table}[ht!]
    \centering
    \begin{tabular}{l c c c c c }
    \hline
    \hline
        \textbf{Parameter} & \hspace{5em} & \textbf{Best-fit} & \hspace{8em}&  & \textbf{68\% limits}\\
        \hline
        $\Omega_bh^2$ & & 0.0225 & & & $0.0226\pm 0.0001$ \\
        $\Omega_ch^2$ & & 0.1195 & & & $0.1196\pm 0.0007$ \\
        $\log(10^{10} \mathcal{A}_s)$ & & 3.052 & & & $3.053\pm0.010$ \\
        $\tau_\mathrm{reio}$ & & 0.05764 & & & $0.0579^{+0.0050}_{-0.0060}$\\
        $100  \theta_s$ & & 1.0406 & & & $1.0407\pm 0.00025$ \\                
        $\delta N$ & & 5.7207 & & & $6.41^{+0.22}_{-1.0}$ \\
         \hline
    \end{tabular}
    \caption{Mean $\pm\ 1 \sigma$ constraints on the inferred cosmological parameters for the effective Starobinsky pre-inflationary model obtained from the joint analysis of P-ACT-LB. For the inferred best-fit value of $\delta N$, we find $N_*\approx54$.}
    \label{tab:cons_table}
\end{table}
\begin{figure}[h!]
    \centering   \includegraphics[width=\columnwidth,height=0.71\columnwidth]{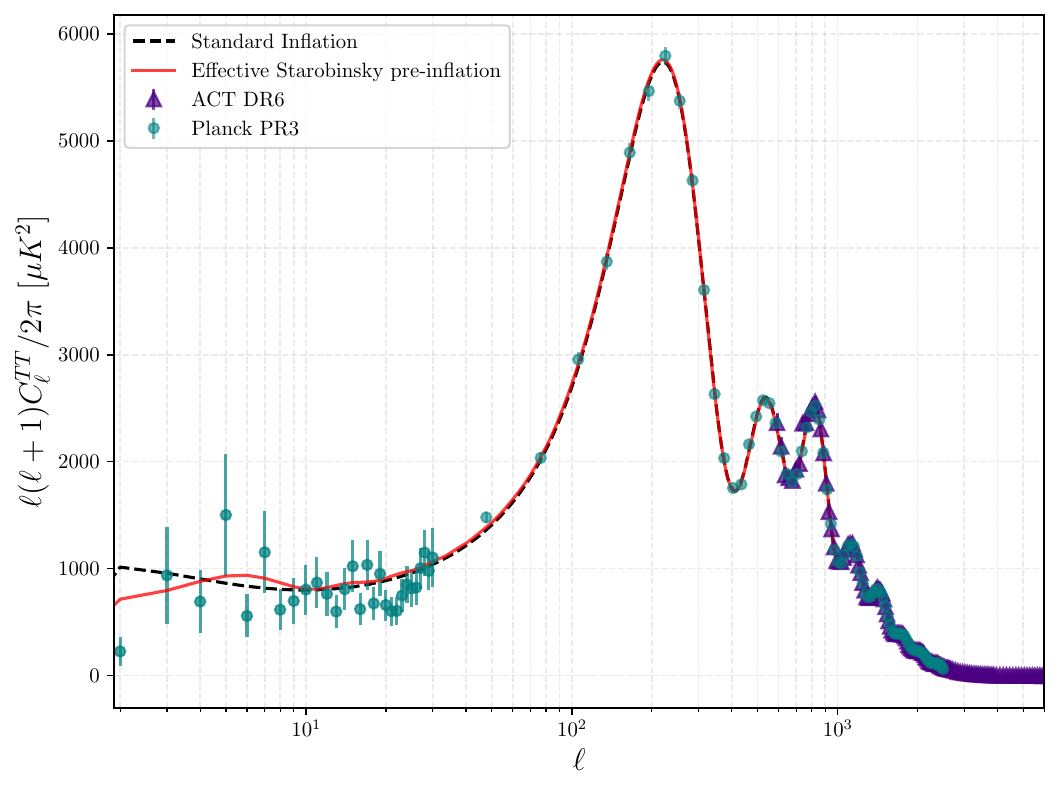}
    \caption{We display the $D_\ell^{TT}$s obtained from Planck PR3 \cite{Planck:2019nip} and ACT DR6 \cite{AtacamaCosmologyTelescope:2025blo}. The black and red curves respectively show the best-fit theoretical $D_\ell^{TT}$s for the standard power law model and the pre-inflationary model using best-fit parameter values from \cref{tab:cons_table}. In the standard power law case, the cosmological parameters are taken from \cite{AtacamaCosmologyTelescope:2025blo} for the P-ACT-LB combination. For the pre-inflationary model, we can see a decrease in power at low $\ell$.
    }
    \label{fig:lambda}
\end{figure}
\begin{figure}[h!]
    \centering       \includegraphics[width=\columnwidth,height=0.71\columnwidth]{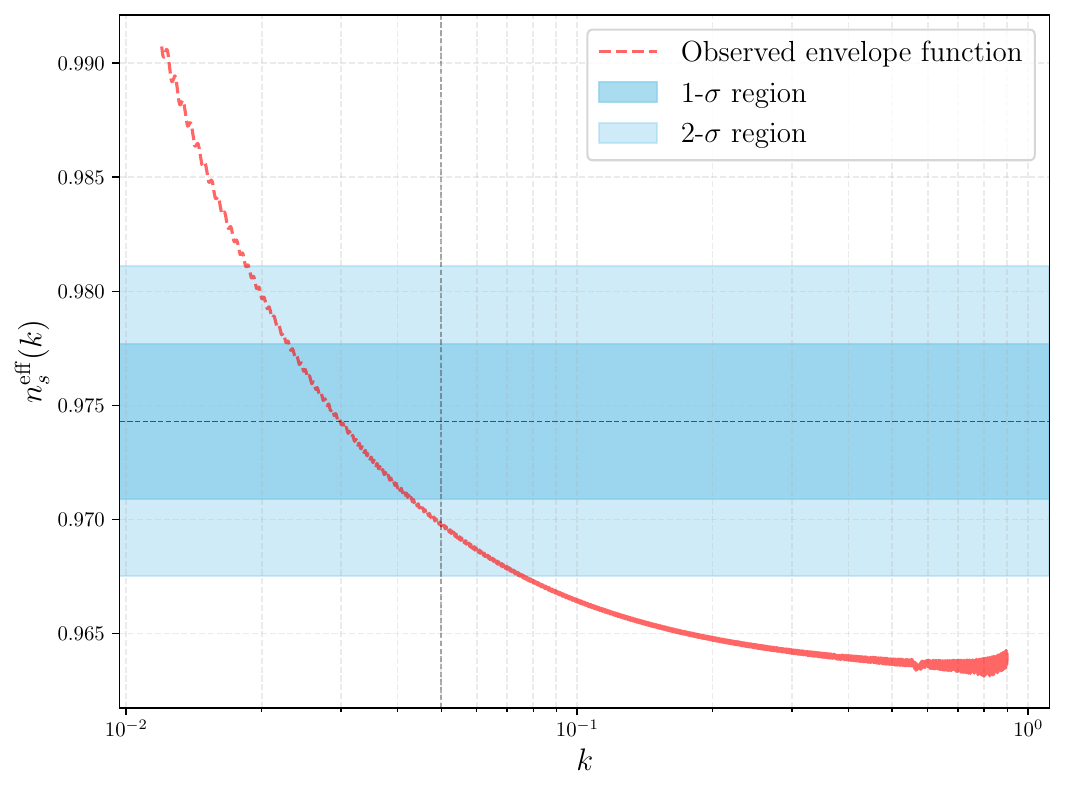}
    \caption{Since $n_s^{\rm eff}$ exhibits decaying oscillations about $n_s^{\rm BD}$, we can obtain the observed value by averaging over the oscillations in an observation window. This is accomplished by a Hilbert transform of the detected signal which extracts the envelope function. The same has been displayed here for the best-fit value $\delta N=5.7207$. The confidence intervals correspond to that of the standard spectral index obtained from the joint analysis of P-ACT-LB \cite{AtacamaCosmologyTelescope:2025blo}.}
    \label{fig:lambda1}
\end{figure}


At the beginning of this work, we highlighted three major issues plaguing Starobinsky model and the inflationary paradigm as a whole: the initial value or fine-tuning problem, the low-$\ell$ power deficit in temperature correlations, and the exclusion of Starobinsky inflation due to the latest ACT results. We have shown that all the three can be resolved by considering quantum corrections to the background dynamics.  We found that the quantum-induced pre-inflationary kination admits a natural initial condition ($\zeta\ll M_P$) which yields the desired inflationary era without the need for fine-tuning. We also found that the low-$\ell$ TT spectrum as well as $n_s^{\rm eff}$ are sensitive to $\delta N$ and, for $\delta N=5.7207$, can be shown to be in reasonable agreement with latest observations (see Figs. \ref{fig:lambda} and \ref{fig:lambda1}).

The best-fit value of $\delta N$ (and consequently $N_*$) is sensitive to the choice of $N_{\rm tot}$. Preliminary investigations hinted that $\delta N$ and $N_*$ increase for larger $N_{\rm tot}$. It can also be shown that $\Delta n_s$ decreases with increase in $\delta N$. So, a choice of larger $N_{\rm tot}$ (provided reheating constraints are not violated for large $N_*$) could provide a larger enhancement in $n_s^{\rm eff}$. Also note that the authors in \cite{Drees:2025ngb} showed that \eqref{eq:nsbd} is only an approximation and by including the leading order corrections, spectral index can be enhanced without the need for any modified inflationary dynamics. Including pre-inflationary kination on top of the modified expression could further enhance $n_s^{\rm eff}$ to even better agreement ACT results.

We conclude, nonetheless, that $N_{\rm tot}=60$ proved to be an excellent choice to test the pre-inflationary kination paradigm in Starobinsky inflation. The resultant best-fit $\delta N=5.7207$ brings $n_s^{\rm eff}$ within 2$\sigma$ of the observed ACT value while preserving standard reheating dynamics in Starobinsky model with $N_*\approx 54$.

    AV thanks Prof. S. Mohanty for fruitful discussions. AV also acknowledges the Council of Scientific \& Industrial Research (CSIR), India for support under the Research Associateship program. SP is partially supported by the DST (Govt. of India) Grant No. SERB/PHY/2021057.

\bibliographystyle{unsrtnat}
\bibliography{refs}

\end{document}


\title{Supplementary Material for ``Eff-ACT-ive Starobinsky pre-inflation"}

\author{Abhijith Ajith}

\author{Hardik Jitendra Kuralkar}


\author{Sukanta Panda}


\author{Archit Vidyarthi}
       

\maketitle 

\section{Obtaining the Perturbative form of Starobinsky action}

\noindent In this section, we outline the derivation to obtain the action (1) in the main text as performed in [12].
Starting with the Starobinsky model,
\begin{equation}\label{eq:starobinsky-sm-action}
    S=\int d^4x \sqrt{-g}\left[\frac{M_P^2}{2}\left(R+\frac{\alpha}{2M_P^2}R^2\right)\right],
\end{equation}
one can replace the quadratic curvature term by an auxiliary scalar using the Lagrange multiplier method,
\begin{equation}\label{eq:starobinsky-jordan-lagrange}
    S=\int d^4x \sqrt{-g}\left[\frac{M_P^2}{2}\left(1+\frac{\alpha\chi^2}{M_P^2}\right)R-\frac{\alpha\chi^4}{4}\right].
\end{equation}
Though we have separated the longitudinal scalar mode, in the form of field $\chi$, and expressed the action in linear order of $R$, $\chi$ is still non-dynamical. In [12], we obtain a kinetic term for $\chi$ without sacrificing the non-minimal coupling. To do so, we first expand gravitational perturbations about a Minkowski background $g_{\mu\nu}=\eta_{\mu\nu}+\kappa h_{\mu\nu}$, where $\kappa=2/ M_P$. This choice of background is relevant for high-energy processes and best suited for probes into deep horizon physics in expanding backgrounds. Now, we can write, up to quadratic order in $h_{\mu\nu}$,
\begin{equation}\label{eq:quadratic-EH}
    \sqrt{-g}\frac{M_P^2}{2}R=M_P(\partial_\mu\partial_\nu h^{\mu\nu}-\Box h)-\partial_\mu h^{\mu\nu}\partial_\nu h+\frac{1}{2}\partial_\mu h\partial^\mu h 
    -\frac{1}{2}\partial_\rho h^{\mu\nu}\partial^\rho h_{\mu\nu}+\partial_\mu h^{\mu\nu}\partial^\rho h_{\rho\nu}+\frac{1}{M_P}\mathcal{O}(h^3).
\end{equation}
Substituting \eqref{eq:quadratic-EH} in \eqref{eq:starobinsky-jordan-lagrange}, we come across cross kinetic mixing terms of the form $\chi^2 \partial\partial h$ but no independent kinetic terms for $\chi$. It can be obtained by performing the linear translation  $h_{\mu\nu}\to h_{\mu\nu}-\eta_{\mu\nu}n\kappa\chi^2$, where for $n=\alpha/4$ all cross kinetic terms cancel out. Note that this is simply a linearized Weyl transformation about a Minkowski background provided that $\chi/M_P\ll1$. Then, from \eqref{eq:starobinsky-jordan-lagrange}, we have,
\begin{align}
    S=\int d^4x &\left[\left(1+\frac{\alpha\chi^2}{M_P^2}\right)\left(-\partial_\mu h^{\mu\nu}\partial_\nu h+\frac{1}{2}\partial_\mu h\partial^\mu h -\frac{1}{2}\partial_\rho h^{\mu\nu}\partial^\rho h_{\mu\nu}+\partial_\mu h^{\mu\nu}\partial^\rho h_{\rho\nu}\right)\right.\nonumber\\
    &\left.-\frac{3\alpha^2}{4M_P^2}\partial_\mu\chi^2\partial^\mu\chi^2+\frac{\alpha^2}{M_P^3}\chi^2(\partial_\mu h^{\mu\nu}\partial_\nu\chi^2-\partial_\mu h\partial^\mu\chi^2)+\frac{3\alpha^3}{4M_P^4}\chi^2\partial_\mu\chi^2\partial^\mu\chi^2-\frac{\alpha\chi^4}{4}\right],
\end{align}
up to quadratic order in $h_{\mu\nu}$ (we have omitted relevant corrections from $\mathcal{O}(h^3)$ terms in \eqref{eq:quadratic-EH} for typographical ease). Having obtained the kinetic term for $\chi^2$, we proceed by canonicalizing the scalar DOF using the transformation,
\begin{equation}\label{eq:canonicalization-chi-zeta}
    \sqrt{\frac{3}{2}}\frac{\alpha}{M_P}\chi^2=\zeta,
\end{equation}
after which we arrive at the action,
\begin{align}\label{eq:quadratic-action-zeta}
    S=\int d^4x &\left[\left(1+\sqrt{\frac{2}{3}}\frac{\zeta}{M_P}\right)\left(-\partial_\mu h^{\mu\nu}\partial_\nu h+\frac{1}{2}\partial_\mu h\partial^\mu h-\frac{1}{2}\partial_\rho h^{\mu\nu}\partial^\rho h_{\mu\nu}+\partial_\mu h^{\mu\nu}\partial^\rho h_{\rho\nu}\right)\right.\nonumber\\
    &\left.-\frac{1}{2}\partial_\mu\zeta\partial^\mu\zeta-\frac{M_P^2}{6\alpha}\zeta^2+\frac{2}{3}\frac{\zeta}{M_P}(\partial_\mu h^{\mu\nu}\partial_\nu\zeta-\partial_\mu h\partial^\mu\zeta)+\frac{1}{\sqrt{6}}\frac{\zeta}{M_P}\partial_\mu\zeta\partial^\mu\zeta\right],
\end{align}
where $\zeta$ is identified as the scalaron which is the scalar inflaton associated with Starobinsky inflation with mass,
\begin{equation}
    M_\zeta\equiv \frac{M_P}{\sqrt{3\alpha}}\approx(3.173\pm0.022)\times10^{13}\text{GeV}.
\end{equation}
Note that this action is only valid in scenarios where the background curvature nearly flat and the scalaron magnitude is small. In the main text, we justify these conditions by working deep inside the horizon and assuming a pre-inflationary era where the scalaron-driven inflation has not yet begun.

\section{Integrating out graviton modes}

\noindent Starting with the action \eqref{eq:quadratic-action-zeta}, we want to integrate out the graviton modes to obtain a graviton-loop corrected action for the scalaron modes $\zeta$. From the standard effective action approach, we know for a given theory,
\begin{equation}
    S_\phi=\int d^4x\ \frac{1}{2} \psi(\Box-m^2)\psi,
\end{equation}
upon integrating out the $\psi$ modes, we obtain quantum corrections of the form: $m^4$, i.e. the effective mass squared. From \eqref{eq:quadratic-action-zeta}, we can see that the graviton part of the action can also be expressed in this form (the non-trivial kinetic part can be fixed by gauge-fixing). The graviton effective mass term, then, turns out to be of the form: $h(\Box\zeta)h$, where the quantity inside the brackets provides an effective mass to the gravitons through non-linear kinetic interaction with $\zeta$. Following the effective action prescription then yields,
\begin{equation}
    S_{\rm eff}\approx\int d^4x \left[-\frac{1}{2}\partial_\mu\zeta\partial^\mu\zeta-\frac{1}{2}M_\zeta^2\zeta^2+\frac{1}{\sqrt{6}}\frac{\zeta}{M_P}\partial_\mu\zeta\partial^\mu\zeta+\frac{\beta}{M_P^2}\Box\zeta\Box\zeta\right],
\end{equation}
for the remaining scalaron DOF. 

\section{Decay of unstable Lee-Wick field into stable scalaron modes}
According to the Ostrogradsky theorem, the higher derivative term can introduce ghost modes in the system. These unstable modes can decay into stable modes via decay vertices that can be found through available interaction terms as shown for the case of Lee-Wick fields in [15]. Assuming $\beta\ll M_P^2/M_\zeta^2$ in the quantum corrected action above, the momentum-space propagator for $\zeta$ looks like,
\begin{equation}
    \frac{i}{p^2-\frac{\beta p^4}{M_P^2}+M_\zeta^2}\approx\left(\frac{i}{p^2}+\frac{i}{-p^2+\frac{M_P^2}{\beta}}\right).
\end{equation}
It is clear that due to the quartic derivative term, a ghost DOF with mass $M_P/\sqrt{\beta}$ is present in the system in addition to the non-ghost scalar. Rewriting the action in terms of stable field $\hat{\zeta}$ and unstable field $\tilde{\zeta}$ (following the analysis done in [15]),
\begin{align}
    S_{eff}\approx\int d^4x &\left[-\frac{1}{2}\partial_\mu\hat{\zeta}\partial^\mu\hat{\zeta}+\frac{1}{2}\partial_\mu\tilde{\zeta}\partial^\mu\tilde{\zeta}-\frac{M_P^2}{2\beta}\tilde{\zeta}^2-\frac{1}{2}M_\zeta^2\hat{\zeta}^2+\frac{1}{\sqrt{6}}\frac{\hat{\zeta}}{M_P}\partial_\mu\hat{\zeta}\partial^\mu\hat{\zeta}-\sqrt{\frac{2}{3}}\frac{\hat{\zeta}}{M_P}\partial_\mu\hat{\zeta}\partial^\mu\tilde{\zeta}-\frac{1}{\sqrt{6}}\frac{\tilde{\zeta}}{M_P}\partial_\mu\hat{\zeta}\partial^\mu\hat{\zeta}\right.\nonumber\\
    &\left.+\sqrt{\frac{2}{3}}\frac{\tilde{\zeta}}{M_P}\partial_\mu\tilde{\zeta}\partial^\mu\hat{\zeta}+\frac{1}{\sqrt{6}}\frac{\hat{\zeta}}{M_P}\partial_\mu\tilde{\zeta}\partial^\mu\tilde{\zeta}-\frac{1}{\sqrt{6}}\frac{\tilde{\zeta}}{M_P}\partial_\mu\tilde{\zeta}\partial^\mu\tilde{\zeta}\right],
\end{align}
we see both decay as well as annihilation channels for $\tilde{\zeta}$ to produce $\hat{\zeta}$. Considering only the $\tilde{\zeta}\to\hat{\zeta}\hat{\zeta}$ decay channels, we find that the contributing vertices are,
\begin{equation}
    -\sqrt{\frac{2}{3}}\frac{\hat{\zeta}}{M_P}\partial_\mu\hat{\zeta}\partial^\mu\tilde{\zeta}-\frac{1}{\sqrt{6}}\frac{\tilde{\zeta}}{M_P}\partial_\mu\hat{\zeta}\partial^\mu\hat{\zeta}.
\end{equation}
Proceeding to the momentum space, the momenta of the produced fields is related to the mass of the decaying field via $2k=M_P/\sqrt{\beta}$. Then, using the standard relations, we find the decay width of $\tilde\zeta$ to be,
\begin{equation}
    |\Gamma|\sim\frac{k^4}{M_P^2}\times\frac{\sqrt{\beta}}{M_P}=\frac{M_P}{\beta^{3/2}},
\end{equation}
up to some constant coefficients that depend on the coupling vertex coefficients. We only consider the magnitude of the decay width here because for Lee-Wick fields, the decay width appears with the opposite sign than for stable particles (see [15] for an in-depth discussion).

\section{Dynamical Stability Analysis for Kination-Inflation Transition}\label{sec:dynamical}
\noindent Given the action,
 \begin{equation}
    S\approx\int d^4x \sqrt{-g}\left[\frac{M_P^2}{2}R-\frac{1}{2}\left(1-\sqrt{\frac{2}{3}}\frac{\zeta}{M_P}\right)\partial_\mu\zeta\partial^\mu\zeta-\frac{1}{2}M_\zeta^2\zeta^2\right],
\end{equation}
we define the following dynamical variables,
\begin{equation}
    x^2=\frac{\Dot{\zeta}^2}{6H^2M_P^2}\left(1-\sqrt{\frac{2}{3}}\frac{\zeta}{M_P}\right),\
    y^2=\frac{M_\zeta^2\zeta^2}{6H^2M_P^2},\
    z^2=\frac{H^2}{M_P^2},
\end{equation}
with constraints,
\begin{equation}\label{eq:stability-constraints}
    x^2+y^2=1,\quad 0\leq x^2, y^2\leq1,\quad yz\leq\frac{M_\zeta}{2M_P}.
\end{equation}
Note that the last constraint is equivalent to $\zeta\leq\sqrt{3/2}M_P$ and the condition $x^2\geq0$ ensures that kinetic energy density remains positive throughout its evolution. Next, writing only the time dependent part of $\zeta$'s equation of motion since the spatial part either decays (sub-Hubble modes) or oscillates slowly (super-Hubble modes),
\begin{equation}\label{eq:eom-zeta}
    (\Ddot{\zeta}+3H\Dot{\zeta})\left(1-\sqrt{\frac{2}{3}}\frac{\zeta}{M_P}\right)-\frac{\Dot{\zeta}^2}{\sqrt{6}M_P}+M_\zeta^2\zeta=0.
\end{equation}
Similarly, the Friedmann equations can be found to be,
\begin{align}
    H^2&=\frac{1}{3M_P^2}\left[\frac{\Dot{\zeta}^2}{2}\left(1-\sqrt{\frac{2}{3}}\frac{\zeta}{M_P}\right)+\frac{1}{2}M_\zeta^2\zeta^2\right],\\
    \Dot{H}&=\frac{-\Dot{\zeta}^2}{2M_P^2}\left(1-\sqrt{\frac{2}{3}}\frac{\zeta}{M_P}\right).
\end{align}
With these, we can write,
\begin{align}
    \frac{dx}{dN}&=-3x-\frac{y}{z}\frac{M_\zeta}{M_P}\left(1-2yz\frac{M_P}{M_\zeta}\right)^{-1/2}+3x^3,\nonumber\\
    \frac{dy}{dN}&=3x^2y+\frac{x}{z}\frac{M_\zeta}{M_P}\left(1-2yz\frac{M_P}{M_\zeta}\right)^{-1/2},\nonumber\\
    \frac{dz}{dN}&=-3x^2z,
\end{align}
where $dN=Hdt$ is the number of e-folds. Using these, we can proceed to analyze stability of the four regimes: kination ($x^2=1$, $y^2=0$), radiation-like ($x^2=2/3$, $y^2=1/3$), matter-like ($x^2=1/2$, $y^2=1/2$), and inflation ($x^2=0$, $y^2=1$). Results have been summarized in Table \ref{table1} and indicate clearly that inflation directly follows from the kination era. Note that near the start of inflation, we use the relation for the energy density of the universe during potential-domination,
\begin{equation}
    \rho_\zeta=\frac{3}{4}M_\zeta^2M_P^2=3M_P^2H_f^2,
\end{equation}
where $H_f\approx6.5\times10^{-6}M_P$ using known value of $M_\zeta\approx1.3\times10^{-5}M_P$. $H_f$ has been used in the main text as a boundary condition when studying the background evolution of $\zeta$ during the kination phase.
\begin{table}[htb]\label{table1}
\begin{ruledtabular}
\begin{tabular}{c  c  c}
$x$, $y$ & $\frac{dx}{dN}$, $\frac{dy}{dN}$, $\frac{dz}{dN}$ & Behavior \\
\hline
$x^2=1$, $y^2=0$ & $\frac{dx}{dN}=0$, $\frac{dy}{dN}>0$, $\frac{dz}{dN}<0$ & Saddle Point \\
$x^2=2/3$, $y^2=1/3$ & $\frac{dx}{dN}<0$, $\frac{dy}{dN}>0$, $\frac{dz}{dN}<0$ & - \\
$x^2=1/2$, $y^2=1/2$ & $\frac{dx}{dN}<0$, $\frac{dy}{dN}>0$, $\frac{dz}{dN}<0$ & - \\
$x^2=0$, $y^2=1$ & $\frac{dx}{dN}\to-\infty$, $\frac{dy}{dN}=0$, $\frac{dz}{dN}=0$ & Attractor \\
\end{tabular}
\end{ruledtabular}
\caption{Pre-inflationary evolution postulated in Section III has been validated here. The pre-inflationary kination era acts a transient epoch leading to the inflationary attractor. Note that inflation appears as a stable attractor here because of the nature of dynamical fields in the system.}
\end{table}

\section{Power Spectrum modification due to Pre-inflationary Kination}
Note that thus far, we had used the start of kination as the temporal reference point. However, inflationary calculations typically denote the end of inflation as the reference point such that conformal time $\tau=\int dt/a\to0^{-}$. As such, for all calculations in the pre-inflationary era, one may assume $\tau\to-\infty$ without causing any issues with the analysis. The scale factor across the transition boundary $\tau_f$ varies as,
\begin{align}
    a(\tau)&=a_f\sqrt{\frac{\tau}{\tau_f}},\qquad \text{for}\ \tau<\tau_f,\\
    a(\tau)&=a_f\frac{H_f}{H}\frac{\tau_f}{\tau},\ \ \text{for}\ \tau_f\leq\tau<0^-.
\end{align}
Using the ADM formalism, we can write the metric as,
\begin{equation}
    ds^2=-N^2dt^2+h_{ij}(dx^i+N^idt)(dx^j+N^jdt),
\end{equation}
 where $N$ is called the lapse function and $N^i$ is called the shift function. Working in the comoving gauge, we can choose, $\delta\zeta=0$. Then, scalar perturbations only appear in the metric,
 \begin{equation}
     h_{ij}=a^2e^{2\mathcal{R}}\delta_{ij},
 \end{equation}
 where $\mathcal{R}$ represents scalar curvature perturbations. The action,
 \begin{equation}
    S\approx\int d^4x \sqrt{-g}\left[\frac{M_P^2}{2}R-\frac{1}{2}\left(1-\sqrt{\frac{2}{3}}\frac{\zeta}{M_P}\right)\partial_\mu\zeta\partial^\mu\zeta-\frac{1}{2}M_\zeta^2\zeta^2\right].
\end{equation}
 can be expanded to second order in $\mathcal{R}$ to give,
 \begin{equation}
     S^{(2)}=\frac{1}{2}\int d\tau\ d^3x\ a^3Q\left[\mathcal{R}'^2-(\nabla\mathcal{R})^2\right],
 \end{equation}
 where $Q=K(\zeta)\Dot{\zeta}^2/H^2$ and $K(\zeta)=1-\sqrt{2/3}\zeta/M_P$ is the non-linear modification to the kinetic term. Defining,
 \begin{equation}
     z\equiv a\sqrt{Q}=\frac{a\sqrt{K(\zeta)}\Dot{\zeta}}{H},
 \end{equation}
 we find that the Mukhanov-Sasaki equation during kination becomes,
 \begin{equation}\label{eq:kination-ms}
     \nu_k''+\left(k^2+\frac{1}{4\tau^2}\right)\nu_k=0,
 \end{equation}
 where $\nu_k=z\mathcal{R}$ is the Mukhanov-Sasaki variable, $'$ represents a derivative with respect to $\tau=\int dt/a(t)$ which is the conformal time, and we have assumed that deep within the horizon, $K(\zeta)\sim $ constant due to logarithmic growth of $\zeta$. \eqref{eq:kination-ms} has solutions,
 \begin{equation}
     \nu_k(\tau)=\sqrt{|\tau|}[A_kJ_0(k\tau)+B_kY_0(k\tau)].
 \end{equation}
 This implies that sub-horizon $\mathcal{R}$ oscillates and decays while super-horizon $\mathcal{R}$ varies slowly but isn't constant. We assume that modes that exit the horizon near the pivot scale originated deep within the horizon during kination. Now, from the assumption of Minkowskian vacuum in deep horizon,
 \begin{equation}
     \nu_k \to \frac{1}{\sqrt{2k}} e^{-ik\tau}
 \quad \text{as } k\tau \to -\infty,
 \end{equation}
 implying,
 \begin{equation}
     A_k = \sqrt{\frac{\pi}{4k}}, \quad B_k = iA_k,
 \implies \nu_k(\tau) =\sqrt{\frac{\pi|\tau|}{4k}} \left[ J_0(k\tau) + i Y_0(k\tau) \right].
 \end{equation}
 This behavior is unlike standard Starobinsky inflation where the Mukhanov-Sasaki equation is,
 \begin{equation}
     \mu_k''+\left[k^2-\frac{1}{\tau^2}\left(\chi^2-\frac{1}{4}\right)\right]\mu_k=0,
 \end{equation}
 where $\mu_k=\Bar{z}\Bar{\mathcal{R}}$, $\Bar{z}=a\Dot{\phi}/H$, $\Bar{z}''/\Bar{z}=2/\tau^2$, and $\chi^2\equiv9/4-3\eta+4\epsilon$ where $\epsilon$ and $\eta$ are slow-roll parameters. The general solution to this equation is,
 \begin{equation}
     \mu_k = \sqrt{|\tau|}\left[C_kH^{(1)}_{\chi}(k\tau) + D_k H^{(2)}_{\chi}(k\tau) \right].
 \end{equation}
 Now, asymptotically as $k\tau\to-\infty$,
 \begin{equation}
     H^{(1)}_{3/2}(k\tau) \sim \sqrt{\frac{2}{\pi k|\tau|}} e^{-ik\tau},\quad\text{and}\quad
     H^{(2)}_{3/2}(k\tau) \sim \sqrt{\frac{2}{\pi k|\tau|}} e^{ik\tau},
 \end{equation}
 which gives $ C_k = 1,\ D_k = 0,$ for Bunch-Davies vacuum (asymptotic behavior is modified due to pre-inflationary kination). Alternatively, for $k\tau\to-\infty$, we have,
 \begin{equation}
     \mu_k(\tau) \approx \frac{1}{\sqrt{2k}} \left[ \alpha_k e^{-ik\tau} + \beta_k e^{ik\tau} \right],
 \end{equation}
 where $\alpha_k$ and $\beta_k$ are Bogoliubov coefficients such that $|\alpha_k|^2-|\beta_k|^2=1$. Therefore,
 \begin{equation}
     \alpha_k=iD_k,\qquad \beta_k=-iC_k.
 \end{equation}
 The modified scalar power spectrum due to deviations from Bunch-Davies vacuum can be expressed as,
 \begin{equation}\label{eq:power-spectrum-modified}
     \mathcal{P}_s(k)=\mathcal{P}^{BD}_s(k)T(k),
 \end{equation}
 where,
 \begin{align}
     \mathcal{P}^{BD}(k)=&\mathcal{A}_s\left(\frac{k}{k_*}\right)^{n_s^{BD}-1},\\
     \text{and}\quad T(k)=&|\alpha_k+\beta_k|^2=|C_k-D_k|^2,
 \end{align}
 and $A_s\approx2.9\times10^{-9}$ as per Planck 2018 data [10], $k_*$ is the pivot scale at which the observed modes exit the horizon before the end of inflation, and $n_s^{\rm BD}$ is the expected spectral index for a given standard inflation inflationary paradigm. Note that we work with $k_*=0.05\ \text{Mpc}^{-1}$ for which, for Starobinsky inflation, $N_*\approx54$ and $n_s^{\rm BD}=0.9649$.
 
 This expression is in agreement with the result obtained in [16]. We demand continuity of the wavefunction of gauge invariant inflaton fluctuations and its time derivative across the transition time $\tau = \tau_f$. The physical wavefunctions can be obtained as,
 \begin{equation}
     \delta\zeta= a^{-1}\nu_k,\quad \text{and}\quad \delta\phi=a^{-1}\mu_k,
 \end{equation}
 and the corresponding matching conditions are,
 \begin{align}
     \delta\zeta(\tau_f) &= \delta\phi(\tau_f),\\
    {\delta\zeta'}(\tau_f) &= {\delta\phi'}(\tau_f).
 \end{align}
The scale factor across the transition boundary varies as,
\begin{align}
    a(\tau)&=a_f\sqrt{\frac{\tau}{\tau_f}},\qquad \text{for}\ \tau<\tau_f,\\
    a(\tau)&=a_f\frac{H_f}{H}\frac{\tau_f}{\tau},\ \ \text{for}\ \tau_f\leq\tau<0^-
\end{align}
 where subscript $f$ is used to indicate values evaluated at the transition boundary, i.e. at the start of inflation. The matching conditions give us,
 \begin{align}
     \frac{1}{\sqrt{2k\tau_f}}e^{-ik\tau_f}&= C_k H^{(1)}_{\chi} + D_k H^{(2)}_{\chi},\\
     -\left(\frac{1}{2\tau_f}+ik\right)\frac{1}{\sqrt{2k\tau_f}}e^{-ik\tau_f}&= C_k H^{\prime(1)}_{\chi} + D_k H^{\prime(2)}_{\chi}.
 \end{align}
 Solving this system of equations, we find,
 \begin{equation}
    T(k)=\left|C_k-D_k\right|^2= \frac{\pi}{2 k|\tau|}\left[(-k\tau)^2\left(J_{\chi}^2\left(k\tau\right)+J_{\chi+1}^2\left(k\tau\right)\right)+2 k\tau\left(\chi+\frac{1}{2}\right) J_{\chi}\left(k\tau\right) J_{\chi+1}\left(k\tau\right)+\left(\chi+\frac{1}{2}\right)^2 J_{\chi}^2\left(k\tau\right)\right]\Bigg|_{\tau_f}.
 \end{equation}
 Assuming that it takes $\delta N$ e-folds for the modes from the start of inflation to exit the horizon,
 \begin{equation}
    \tau_f=\tau_*e^{\delta N}\frac{H_*}{H_f}.
\end{equation}
 The consequent effects on the spectral index can be quantified as,
 \begin{align}
    n_s - 1 &= \frac{d\ln \mathcal{P}_s(k)}{d\ln k},\nonumber\\
    &= \frac{d\ln \mathcal{P}^{\text{BD}}_s(k)}{d\ln k} + \frac{d\ln{T(k)}}{d\ln k},\\
    \implies\Delta n_s&=\frac{d\ln{T(k)}}{d\ln k}.
 \end{align}

\section{Inference of cosmological parameters}
\begin{figure}[h!]
    \centering       \includegraphics[width=0.75\linewidth,height=0.5\linewidth]{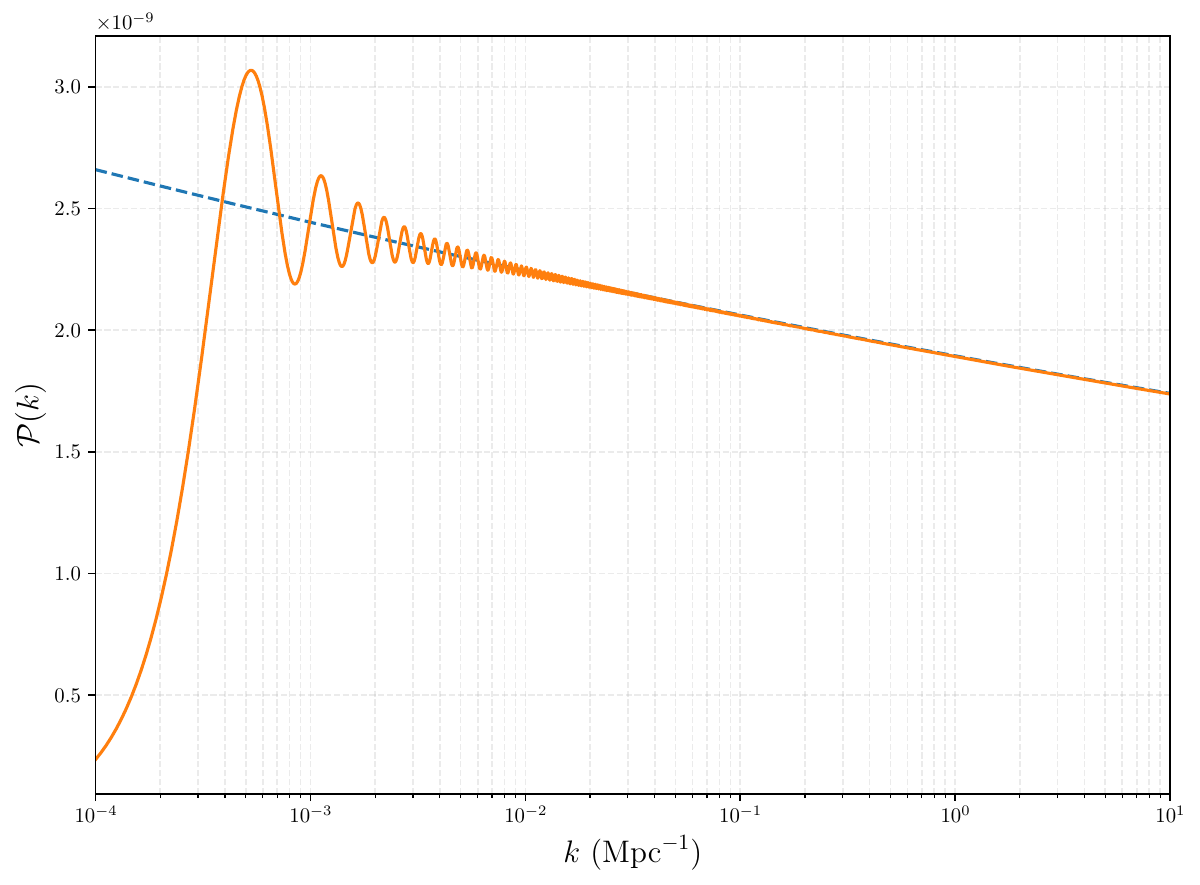}
    \caption{Variation of the primordial scalar power spectrum with $k$. We see that for $k$ near the pivot scale, the modified spectrum shows oscillations about the observed (nearly) scale-invariant power spectrum. The modified amplitude decays for small $k$. We have used the best fit values of the cosmological parameters inferred from the MCMC analysis for obtaining the scalar power spectrum.}
    \label{fig:MPS}
\end{figure}

\begin{figure*}[ht!]
    \centering   
    \includegraphics[width=\textwidth]{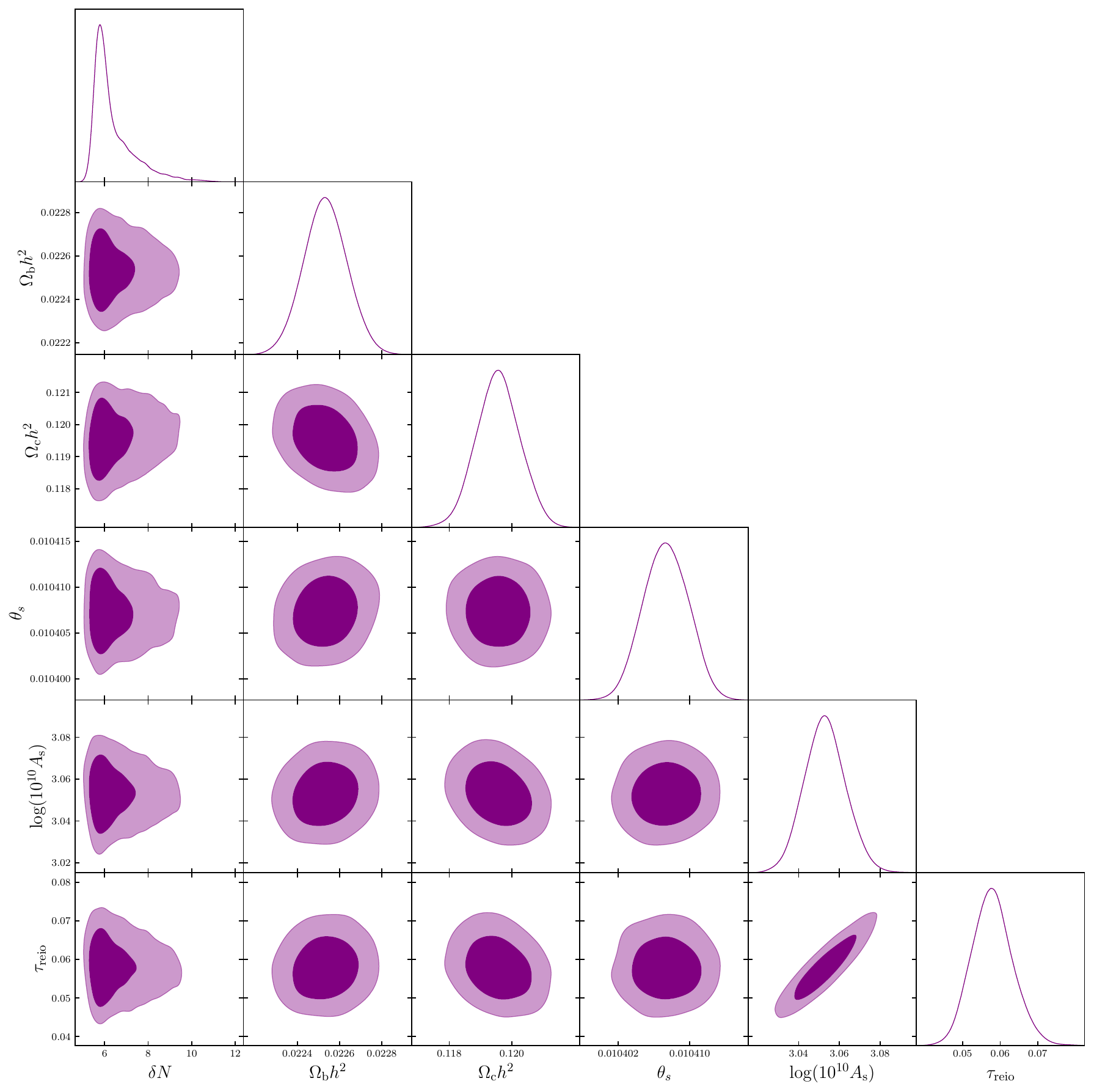}
    \caption{Corner plots of 1D and 2D marginalized posterior distributions for the effective Starobinsky pre-inflationary model, based on CMB from Planck and ACT DR6, lensing from ACT and BAO from DESI DR1. Contours at 68\% ($1\sigma$) and 95\% ($2\sigma$) levels
showing parameter constraints and correlations within the framework.}
    \label{fig:corner}
\end{figure*}

\noindent Here, we display some figures related to the parameter inference obtained using the modified power spectrum that arises in the effective Starobinsky pre-inflationary model. The corner plot of the cosmological parameters that comprise our parameter space is shown in \cref{fig:corner}, while the behavior of the modified power spectrum is showcased in \cref{fig:MPS}

